\documentclass[12pt]{iopart}
\usepackage{graphicx}
\usepackage{bbm}
\begin{document}

\title{Nonclassical Correlation of Polarisation Entangled Photons in a Biexciton-Exciton Cascade}
\author{Sumanta Das and G. S. Agarwal}
\address{Department of Physics, Oklahoma State University,
Stillwater, OK - 74078, USA} \eads{\mailto{sumanta.das@okstate.edu},
\mailto{agirish@okstate.edu}}
\date{\today}

\begin{abstract}
We develop a theoretical model to study the Intensity-Intensity correlation of polarisation entangled photons emitted in a biexciton-exciton cascade. We calculate the degree of correlation and show how polarisation correlations are affected by the presence of  dephasing and energy level splitting of the excitonic states. Our theoretical calculations are in agreement with the recent observation of polarisation dependent Intensity-Intensity correlations from a single semiconductor quantum dot [R. M. Stevenson \textit{et. al.} \textit{Nature} { \bf 439}, 179 (2006)] . Our model can be extended to study polarisation entangled photon emission in coupled quantum dot systems.
\end{abstract}
\maketitle

\section{Introduction}
Polarisation correlations of photons emitted in cascade emission are
 well known phenomenon and numerous theoretical and experimental
studies exist in the literature on this subject since the early days of quantum optics \cite{1,2,3,4,4a,4b}.
Some of the earlier studies were motivated in testing generalised Bell's inequalities \cite{2}, the existence of hidden variables and whether quantum mechanics was a non-local
theory or not \cite{3,4}, following the question raised by Einstein, Podolsky and Rosen \cite{5}. 
 In recent times polarisation
correlated  photon pairs have become important in the field of
quantum information science due to their entangled nature.
 Moreover many applications of quantum information, such as quantum key distribution \cite{9},
efficient optical quantum computing \cite{10}, long distance quantum communication
using quantum repeaters \cite{11} and implementation of quantum telecommunication
schemes \cite{chandelier} require single photon pairs per cycle. This requirement of entangled photon pairs per cycle of excitation could be satisfied by cascade emission from a single
atom or atom like systems like semiconductor quantum dots, provided one gets over the inherent asymmetries. Recently such cascade
emission has been reported for semiconductor quantum dots
\cite{benson,santori,gershoni,stevenson}. It was further seen that
polarisation entanglement of the emitted photon pairs was degraded by the presence of energy
level splitting of the intermediate excitonic states and any
incoherent process that leads to a population transfer between the
two intermediate excitonic states 
\cite{gershoni,stevenson,stnjp}. Moreover dephasing arising due
to interaction of the quantum dot with its solid state environment
can also degrade the entanglement \cite{phon}.
Some recent studies 
have also shown how the fidelity of entanglement depends on excitonic level splitting \cite{hud} and the dynamics of the incoherent dephasing \cite{larque}. Different methods have been proposed to
reduce and control the incoherent dephasing and energy level splitting of the excitonic
states thereby preserving the entanglement in the system
\cite{gershoni,stevenson,stnjp,hud,gerd,steg,avron}. Further, methods to enhance the generated entanglement by coupling the quantum dot to a micro-cavity  have also been proposed \cite{tro,johne}. As quantum dot systems
are of great importance for future applications in quantum
information science, a clear yet simple model for understanding the effects of all these different
decoherence mechanism on the dynamics of the system is required. 
Thus we develop, in
this paper, a simple theoretical model to analytically study the influence of
different decoherence mechanisms and the intermediate state splitting on the generation of polarisation entangled photon pairs in cascade emission. 

\section{Model}
We consider a four level system under going cascade
emission as our model. We show a schematic diagram of such a cascade in figure 1.
The excited state $|i\rangle$ and the intermediate states
$|\alpha\rangle$, $|\beta\rangle$ would correspond to the biexcitonic and optically active
excitonic states respectively in a quantum dot. Further $|j\rangle$ is taken to be
the ground state.  Here $2\gamma = 2( \gamma_{1}+\gamma_{3})$ is the
total spontaneous emission rate of the state $|i\rangle$,
$2\gamma_{2}$, $2\gamma_{4}$ are the spontaneous emission rates
of the states $|\alpha \rangle$ and $|\beta\rangle$ respectively and $2\gamma_{\beta\alpha}$($2\gamma_{\alpha\beta}$) is the incoherent
dephasing rate of the state $|\alpha\rangle$ ($|\beta\rangle$). 
The energy level splitting of the intermediate state is given by $\Delta$.
\begin{figure}[h!]
\begin{center}
\includegraphics[scale = 0.85]{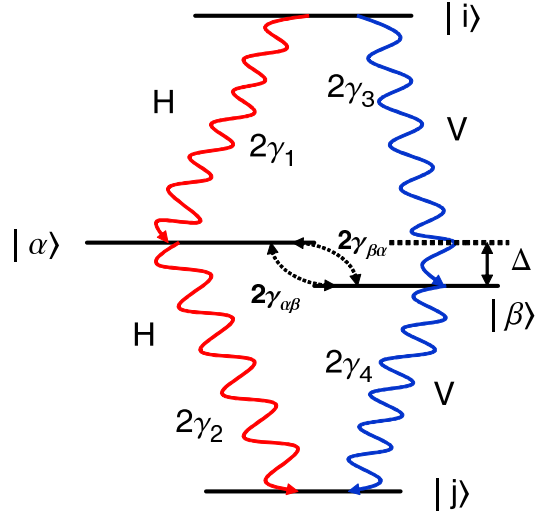}
\caption{Schematic diagram of a four level cascade system. Here H and V refers to horizontally and vertically polarised photon emission. $\Delta$ is the energy level separation of the intermediate states and $\gamma$'s are the spontaneous emission rates given by $\gamma_{k} = 2\omega^{3}_{kl}|\vec{d}_{kl}|^{2}/3\hbar c^3$. The incoherent dephasing rates of the intermediate states are given by $2\gamma_{\alpha\beta}$ and $2\gamma_{\beta\alpha}$ respectively.}
\end{center}
\end{figure}
In this type of four-level cascade scheme there are two decay paths
for the excited state, $|i\rangle \rightarrow |\alpha\rangle
\rightarrow |j\rangle$ and $|i\rangle \rightarrow |\beta\rangle
\rightarrow |j\rangle$.  The generation of entanglement in these scheme is
attributed to the fact that this decay paths can become indistinguishable. The
eigenbasis of this system is formed by the four states
($\{|i\rangle\}, \{ |\alpha\rangle\}, \{|\beta\rangle\},
\{|j\rangle\}$). In this basis the radiative transition from the
excited state generates collinearly polarised photons with linear
polarisations along two orthogonal directions denoted by H
(horizontal) and V (vertical). When the states $|\alpha\rangle$ and
$|\beta\rangle$ are degenerate, the decay paths become 
indistinguishable and we get  a maximally entangled
two photon state \cite{5,benson}
\begin{equation}
\label{1}
|E\rangle = \frac{1}{\sqrt{2}}(|H_{1}H_{2}\rangle + |V_{1}V_{2}\rangle).
\end{equation}
In practical systems like atoms and quantum dots these levels are
usually non degenerate and hence the entanglement of the emitted photon pairs depend completely on the degree of degeneracy and dynamics of these intermediate states.
In  our model we have taken them to be
non-degenerate and study the effect of such intermediate level splitting 
on the correlation of the emitted photon pairs. 
To understand the effect of incoherent dephasing and energy level
splitting of the excitonic state on the dynamics of emitted photon
pairs from the cascade, we need to study the two time
second order  correlations. This is given by,
\begin{eqnarray}
\label{29}
\langle I I \rangle & = &\langle\hat{\epsilon}^{\ast}_{(\theta_{1},\phi_{1})}\cdot \vec{E}^{-}(\vec{r},t)\hat{\epsilon}^{\ast}_{(\theta_{2},\phi_{2})}\cdot \vec{E}^{-}(\vec{r},t+\tau)\nonumber\\
& &:\hat{\epsilon}_{(\theta_{2},\phi_{2})}\cdot \vec{E}^{+}(\vec{r},t+\tau)\hat{\epsilon}_{(\theta_{1},\phi_{1})}\cdot \vec{E}^{+}(\vec{r},t)\rangle.\nonumber\\
\end{eqnarray}
where $\langle I I \rangle$ stands for the two time polarisation angle dependent intensity-intensity correlation $\langle I_{(\theta_{2},\phi_{2})}(\vec{r},t+\tau)I_{(\theta_{1},\phi_{1})}(\vec{r},t)\rangle$.
Further $E^{+}(\vec{r},t)
(E^{-}(\vec{r},t))$ is the positive (negative) frequency part of the
quantized electric field operator at a point $\vec{r}$ in the
far-field zone and $\hat{\epsilon}_{(\theta,\phi)}$ is the polarisation unit
vector of the measured radiation at the detector along any arbitrary
direction given by $(\theta,\phi)$.
\begin{figure}[h!]
\begin{center}
\includegraphics[scale = 0.70]{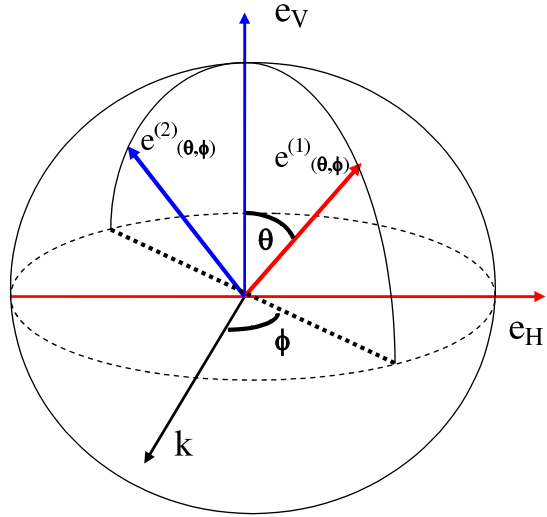}
\caption{Schematic diagram for orientation of the polarisation unit
vectors. Here $\mathbf{e}_{H}$ and $\mathbf{e}_{V}$ corresponds to
horizontal and vertical polarisation respectively.
$\mathbf{e}^{(1)}$ and $\mathbf{e}^{(2)}$ are two arbitrary
orthogonal pair of polarisation unit vectors.\\}
\end{center}
\end{figure}
$\hat{\epsilon}_{(\theta,\phi)}$'s are related to the linear
polarisation unit vectors $\hat{\epsilon}_{H},\hat{\epsilon}_{V}$
(where H stands for horizontal and V for vertical) by ,
\begin{equation}
\label{26}
\left[\begin{array}{c} \hat{\epsilon}^{(1)}_{(\theta,\phi)} \\
\hat{\epsilon}^{(2)}_{(\theta,\phi)}
\end{array}\right] =
\left[\begin{array}{cc} \cos\theta & e^{-i\phi}\sin\theta \\
-e^{i\phi}\sin\theta & \cos\theta \end{array}\right]
\left[\begin{array}{c}\hat{\epsilon}_{H} \\
\hat{\epsilon}_{V}\end{array}
\right]
\end{equation}
and these satisfy the relation
$(\hat{\epsilon}^{(1)}_{(\theta,\phi)}\cdot\hat{\epsilon}^{(2)\ast}_{(\theta,\phi)}
)= 0$. The above relation can be understood as an unitary
transformation between a basis defined by the linear polarisation
unit vectors and a basis defined by $\hat{\epsilon}^{(1)}$ and
$\hat{\epsilon}^{(2)}$. In experimental setup the angles $\theta, \phi$ would correspond to the orientation of the optic axis of a half/quarter wave plate to the direction of propagation of the emitted radiation.  Let us now consider for simplicity that both the levels
$|\alpha\rangle$ and $|\beta\rangle$ in figure. 1 have the same incoherent
dephasing rates i.e. $\gamma_{\alpha\beta} = \gamma_{\beta\alpha}$. 
Further we assume that the spontaneous decay rates of the intermediate levels are also equal. Such assumptions are well justified as they do not influence the dynamics of the system significantly and yet leads to a simplified form of the second order correlation, thereby providing a better understanding of the problem.  Under the above assumptions and for $\phi_{1} = \phi_{2} = 0$ the form of the two-time polarisation angle dependent intensity-intensity correlation is found to be , 
\begin{eqnarray}
\label{34}
\langle I(\theta_{2},t+\tau) I(\theta_{1},t)\rangle& = &\left(\frac{\omega_{0}}{c}\right)^{8}\frac{1}{2r^{4}}\mathcal{D}_{1}^{2}\mathcal{D}_{2}^{2}\langle|i\rangle\langle i|_{t}\rangle\nonumber\\
&\times&\lbrace e^{-2\gamma_{2}\tau}+\cos2\theta_{1}\cos2\theta_{2}e^{-2(\gamma_{2}+2\gamma_{\alpha\beta})\tau}\nonumber\\
&+&\sin2\theta_{1}\sin2\theta_{2}e^{-2(\gamma_{2}+\gamma_{\alpha\beta})\tau}\cos(\Delta\tau)\rbrace
\end{eqnarray}
where $\mathcal{D}_{1} = |\vec{d}_{\alpha i}| = |\vec{d}_{\beta i}|$ and $\mathcal{D}_{2} = |\vec{d}_{ j\alpha }| = |\vec{d}_{j\beta }|$. The above simple form of the second order correlation has been derived to match our theoretical analysis to that of the experiment results \cite{stevenson}. For details of the mathematical analysis leading to the generalised form of the two time intensity-intensity correlation the reader is referred to section 4 of this paper.  One can clearly see from equation (\ref{34}) that the second order correlation is profoundly influenced by both the incoherent dephasing rates as well as the energy level splitting of the intermediate states. Note further, that in the presence of small $\Delta$  this becomes equivalent to the second order correlations measured in ref.\cite{3,4}. Next we define a quantity the degree of correlation $c_{\mu}$ as,
\begin{equation}
c_{\mu} = \frac{\langle I_{\mu}I_{\mu}\rangle -\langle I_{\mu}I_{\mu^{\prime}}\rangle}{\langle I_{\mu}I_{\mu}\rangle +\langle I_{\mu}I_{\mu^{\prime}}\rangle}
\end{equation}
where  $\mu, \mu^{\prime}$ stands for mutually orthogonal polarisation basis. The degree of correlation varies between $+1$ and $-1$, where $+1$ represent perfect correlation ($-1$ for anti-correlation) and 0 represent no polarisation correlation.

\section{Results and Discussion}

\subsection{Effect of excitonic level splitting on the correlation}

In figure. 3(a) we show how the time averaged degree of correlation varies with the basis angle for different values of splitting $\Delta$, of the excitonic levels. Note that here the excitonic level dephasing $\gamma_{\alpha\beta}$ has been taken to be zero. We see that the degree of correlation is independent of the polarisation basis when $\Delta = 0$ and takes a value $c_{\mu} = 1$. This correspond to perfect polarisation correlation among the emitted photons. From the expression of $c_{\mu}$ it is clear that this can happen only when the cross-polarised correlations vanishes and the emitted photons are perfectly co-polarised. One can even see this explicitly from equation (\ref{34}) by putting the values of $\theta_{1}, \theta_{2} = \theta_{1}+\pi/2$ for H-V , D-D$^\prime$ and V-H basis. where H, V, D and D$^\prime$ stands for horizontal, vertical, diagonal and orthodiagonal polarisation basis respectively.
\vspace{0.3in}
\begin{figure}[h!]
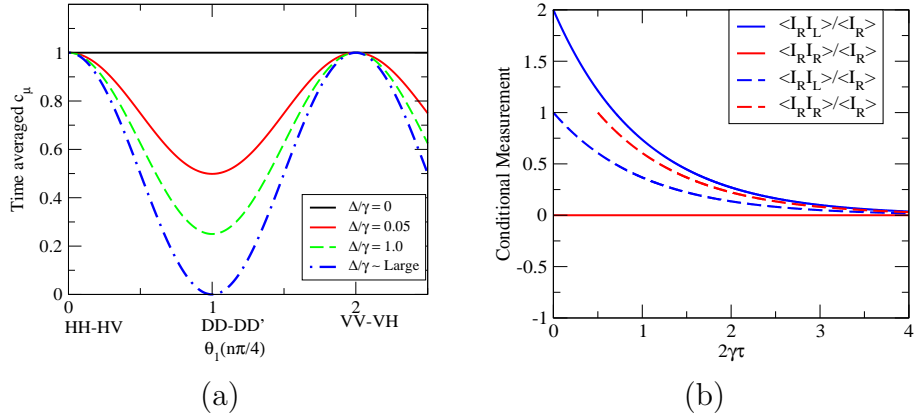

\begin{center}
\begin{tabular}{ccc}
\includegraphics[height = 4.8cm, width = 5.6cm]{c5.eps}& & 
\includegraphics[height = 4.8cm , width = 5.6cm]{corr.eps}\\
(a)& &(b)
\end{tabular}
\caption{(Colour online) (a)Degree of correlation averaged over time
as a function of basis angle, for different excitonic level splitting $\Delta$. 
 H , D, D$^\prime$, V stands for horizontal, diagonal,
orthodiagonal and vertical polarisation basis respectively. Here we have considered zero dephasing of the excitonic states. (b)Conditional measurement of intensity-intensity correlation in the circular basis. The red curve corresponds to co-polarised ($\theta_{1} = \theta_{2} = \pi/4, \phi_{1} = \phi_{2} = -\pi/2$)photons and the blue for cross-polarised ($\theta_{1} = \theta_{2} = \pi/4, \phi_{1} = -\pi/2, \phi_{2} = \pi/2$) ones. The solid curve is for  $\Delta= 0$ and the broken one for $\Delta \sim$ large. Here R and L stands for right and left circular polarisation. The R-R correlation curve in case of large splitting is time shifted for better comparison to the R-L correlation. All parameters are normalized with respect to $\gamma$.}
\end{center}
\end{figure}
Further as the cross-polarised correlations are absent the pair of photons emitted in one excitation cycle can take either of the two paths $|i\rangle \rightarrow |\alpha\rangle \rightarrow |j\rangle$ or $|i\rangle \rightarrow |\beta\rangle \rightarrow |j\rangle$ thus making these paths indistinguishable. As a consequence we do not get  the "Welcher Weg" or which path information thereby making the final state of the emitted photon pair entangled in both the linear and diagonal polarisation basis. The generated entangled states can hence be written as $1/\sqrt{2}(|HH\rangle + |VV\rangle)$ and $1/\sqrt{2}(|DD\rangle + |D'D'\rangle)$ for the rectilinear and diagonal basis respectively. Note further that in this case perfect anti-correlation ($c_{\mu}$ = -1) is expected for measurement in the circularly polarised basis with the entangled state given by $1/\sqrt{2}(|RL\rangle + |LR\rangle)$. Thus one should get perfectly cross-polarised photons as the co-polarised correlations vanish in this basis. This is exactly what we get from the general expression of equation (\ref{34}) [see section 4, equation (\ref{32})] and is shown by the solid curves in figure. 3(b).
Further in figure. 3(a) we see that the degree of correlation is practically independent of the excitonic level splitting $\Delta$ in the rectilinear basis. As we change our polarisation basis the effect of $\Delta$ becomes significant. In the diagonal basis for example with the increase in level splitting the degree of correlation gradually decreases and eventually vanishes. In presence of $\Delta$, the cross-polarisation does not vanish and we have a which path information for the emitted photons when we measure the second order correlations, thus destroying any entanglement in the system. The behaviour of the correlations in the circular basis in presence of large excitonic level splitting is shown by the broken curves in figure. 3(b). One can clearly see that there is no polarisation correlation at all for large $\Delta$. 
The sinusoidal behaviour of $c_{\mu}$ for non zero value of $\Delta$ as seen in figure. 3(a)  is in agreement with the classical linear polarisation correlation behaviour. Note that our theoretical results are  in agreement to experimentally observed data \cite{stevenson}. 

\noindent{}It should be noted that in our analysis we have concentrated on the calculation of the  quantum correlation $c_{\mu}$. This also was measured in the experiment of Stevenson \textit{et. al.} We have not examined measures of entanglement like concurrence. This is because if $\Delta$ -the intermediate state exciton splitting is nonzero then horizontal and vertical photons have different frequencies-which amounts to saying that we have for nonzero $\Delta$ quantum states which are characterized by two different parameters and measures of entanglement in such situations do not exist.

\subsection{ Effect of decoherence  on the correlation}
In figures. 4 (a) and (b) we show how the incoherent dephasing of the intermediate excitonic states affect the time averaged degree of correlations $c_{\mu}$ when excitonic states are non-degenerate ($\Delta \neq 0$) and degenerate ($\Delta = 0$) respectively. Note that here we have assumed that both the intermediate states have same dephasing rates. One can clearly see that the affect is different for different measurement basis. The degree of polarisation correlation for example in the rectilinear basis decreases with increasing dephasing irrespective of whether the excitonic states are non-degenerate or degenerate.
\vspace{0.3in}
\begin{figure}[h!]
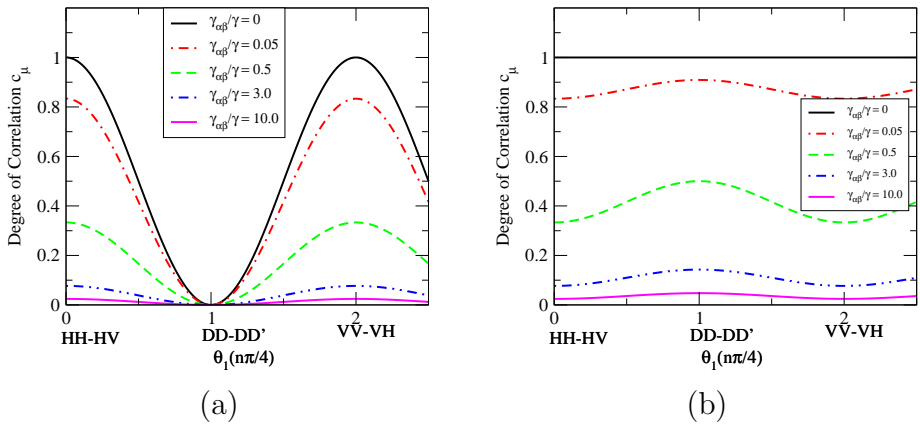

\begin{center}
\begin{tabular}{ccc}
\scalebox{0.4}{\includegraphics{c1.eps}}& & 
\scalebox{0.4}{\includegraphics{c2.eps}}\\
(a)& &(b)
\end{tabular}
\caption{(Colour online) (a)Degree of correlation averaged over time
as a function of basis angle, for large $\Delta$ and different
incoherent dephasing rates $\gamma_{\alpha\beta}$ 
of the intermediate level. Here we have
assumed that both the intermediate states dephase at same rate i.e $\gamma_{\alpha\beta} = \gamma_{\beta\alpha}$. H , D, D$^\prime$, V stands for horizontal, diagonal,
orthodiagonal and vertical basis respectively. (b) Same as (a)
for $\Delta = 0$. }
\end{center}
\end{figure}  
 For large dephasing rates the emitted photon pairs become almost un-correlated in their polarisation. This is attributed to the presence of significant cross-polarised correlation for large dephasing rates of the intermediate states.
The incoherent dephasing of the intermediate levels causes a incoherent population transfer among the states $|\alpha\rangle$ and $|\beta\rangle$ thereby allowing the second photon to be emitted with orthogonal polarisation to the first one. In the diagonal basis on the other hand the dephasing does not affect the correlation at all for large $\Delta$ but significantly decreases the correlation when $\Delta = 0$ for large dephasing rates. 
\begin{figure}
\begin{center}
\includegraphics[height = 4.5cm , width =5.5cm]{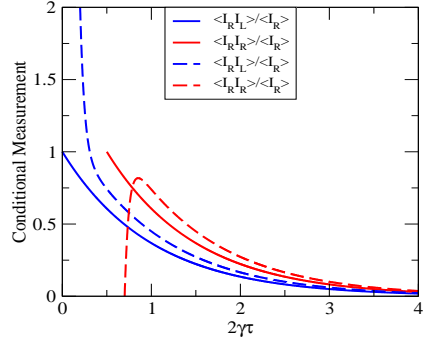}
\caption{ (Colour online) Conditional measurement of intensity-intensity correlation in the circular basis for incoherent dephasing $\gamma_{\alpha\beta}/\gamma = 10$. The red curve corresponds to co-polarised ($\theta_{1} = \theta_{2} = \pi/4, \phi_{1} = \phi_{2} = -\pi/2$)photons and the blue for cross-polarised ($\theta_{1} = \theta_{2} = \pi/4, \phi_{1} = -\pi/2, \phi_{2} = \pi/2$) ones. The solid curve is for $\Delta \sim$ large and broken one for $\Delta = 0$. Here R and L stands for right and left circular polarisation. The R-R correlation curve are time shifted for better comparison to the R-L correlation. }
\end{center}
\end{figure}
So we see that in diagonal basis even when the intermediate levels are degenerate we can still have significant cross-correlation if there is some incoherent relaxation process by which they can get coupled. This in turn spoils the quantum correlation in the system as can be seen clearly from figure. 4(b). In figure. 5 we show how the correlations behave in the circular basis in presence of large dephasing rate ($\gamma_{\alpha\beta}/\gamma = 10$) for both non-degenerate and degenerate intermediate states. We find that in the circular basis decoherence arising due to the incoherent dephasing does not affect the degree of correlation of the emitted photons when $\Delta$ is large.  Further we find that for degenerate intermediate levels, even though the degree of correlation $c_{\mu} = -1$ for zero time delay, it vanishes at all later time in presence of the large dephasing. Thus the decoherence makes the perfectly anti-correlated photons completely uncorrelated. The incoherent relaxation process discussed by us here are practically present in the biexcitonic-excitonic cascade in quantum dots \cite{phon, larque}. Thus we have shown by a simple model how the decoherence arising due to such incoherent processes would strongly affect the quantum correlations. 

\section{Detail derivation of the Intensity-Intensity correlation}
Our model consist of a biexcitonic state and two excitonic states labelled as $|i\rangle$ and $|\alpha\rangle, |\beta\rangle$ respectively. The equilibrium state is labelled as $|j\rangle$.  The biexcitonic state decays by emission of either a horizontally(H) polarised photon $(|i\rangle\rightarrow|\alpha\rangle)$ or a vertically(V) polarised photon $(|i\rangle\rightarrow|\beta\rangle)$\cite{benson,santori}. The excitonic state $|\alpha\rangle(|\beta\rangle)$ decays to the equilibrium state $|j\rangle$ by emission of a H(V)- polarised photon. The two excitonic states have a energy difference of $\hbar\Delta$. Note that the splitting of the excitonic state in quantum dots arises due to anisotropic electron-hole exchange interactions \cite{gam,hom}.  Figure (1) show a schematic diagram of our model. The eigenbasis of this system is formed by the four states ($\{|i\rangle\}, \{ |\alpha\rangle\}, \{|\beta\rangle\},\{|j\rangle\}$). In this basis the unperturbed Hamiltonian $\mathcal{H}$ is given by, 
\begin{eqnarray}
\label{3}
\mathcal{H}& = &\sum_{k} \hbar\omega_{k}|k\rangle\langle k| 
\end{eqnarray}
Where $\hbar\omega_{k}$ is the energy of the four levels ($k =i,\alpha,\beta, j$). Note that this kind of energy level scheme has been extensively used to study the cascade emission in quantum dots \cite{gershoni,stnjp,phon,hud,larque}. The spontaneous emission and dephasing effects in the system are incorporated via a master equation technique \cite{gsab} under the Born, Markov and rotating wave approximations and is given by,
\begin{eqnarray}
\label{3a}
\mathcal{L}\rho& = &-\gamma\lbrace S_{ii}, \rho \rbrace-\gamma_{2}\lbrace S_{\alpha\alpha}, \rho \rbrace-\gamma_{4}\lbrace S_{\beta\beta}, \rho \rbrace-\gamma_{\beta\alpha}\lbrace S_{\alpha\alpha}, \rho \rbrace\nonumber\\
&- & \gamma_{\alpha\beta}\lbrace S_{\beta\beta}, \rho \rbrace +2(\gamma_{1}\rho_{ii}S_{\alpha\alpha}+\gamma_{3}\rho_{ii}S_{\beta\beta}+\gamma_{2}\rho_{\alpha\alpha}S_{jj}\nonumber\\
&+&\gamma_{4}\rho_{\beta\beta}S_{jj})+2\left(\gamma_{\beta\alpha}\rho_{\alpha\alpha}S_{\beta\beta}+\gamma_{\alpha\beta}\rho_{\beta\beta}S_{\alpha\alpha}\right).
\end{eqnarray}
Here $2\gamma =2( \gamma_{1}+\gamma_{3})$ is the
total spontaneous emission rate of the biexcitonic state $|i\rangle$,
$2\gamma_{2}$( $2\gamma_{4}$) and $2\gamma_{\beta\alpha}$($2\gamma_{\alpha\beta}$) are the spontaneous emission rate and incoherent
dephasing rate of the excitonic state $|\alpha \rangle$($|\beta\rangle$) (see Fig. 1). Such incoherent dephasing arises in quantum dots due to it's interaction with the solid-state environment (in form of spin flip processes or phonon scattering) \cite{phon}. The curly bracket $\{.. , ..\}$ stands for the anti-commutator and 
$S_{kl} = |k\rangle\langle l|$($S^{\dagger}_{kl} = |l\rangle\langle k|$ ) is the atomic lowering (raising) operator which follow the simple angular momentum commutation relations.
To study the dynamical evolution of this four-level cascade system we
solve for the time evolution of the density
operator which is given by,
\begin{equation}
\label{2}
\frac{\partial \rho}{\partial t} = -\frac{i}{\hbar}[\mathcal{H}, \rho] + \mathcal{L}\rho ,\
\end{equation}
We will set the energy of the state $|j\rangle$ equal to zero henceforth. On substituting
equations (\ref{3}) and (\ref{3a}) in (\ref{2}) and solving for the population terms
we get,
\begin{eqnarray}
\label{6}
p_{i}(t) &= &e^{-2\gamma t}p_{i}(0);\qquad p_{i} = \rho_{ii}-R/2\gamma\nonumber\\
\rho_{\alpha\beta}(t) &= &e^{-(a_{0}-i\Delta)t}\rho_{\alpha\beta}(0);\nonumber\\
\rho_{\alpha\alpha}(t)& = &e^{-a_{0}t}\left(\cosh(At)+\frac{\Gamma_{a}}{A}\sinh(At)\right)\rho_{\alpha\alpha}(0)\nonumber\\
&  &+ 2e^{-a_{0}t}\frac{\gamma_{\alpha\beta}}{A}\sinh(At)\rho_{\beta\beta}(0)\nonumber\\
& &+ \frac{R}{2\gamma}C(t)+ p_{i}(0)e^{-2\gamma t}D(t);\nonumber\\
\rho_{\beta\beta}(t)& = &e^{-a_{0}t}\left(\cosh(At)-\frac{\Gamma_{a}}{A}\sinh(At)\right)\rho_{\beta\beta}(0)\nonumber\\
& &+ 2e^{-a_{0}t}\frac{\gamma_{\beta\alpha}}{A}\sinh(At)\rho_{\alpha\alpha}(0)\nonumber\\
& &+ \frac{R}{2\gamma}F(t)+ p_{i}(0)e^{-2\gamma t}K(t).
\end{eqnarray}
Here R signifies a constant feeding of population into the state
$|i\rangle$ from some arbitrary state $|n\rangle$. Note that in our
model we are only concerned with the dynamics of the cascade decay
once the upper level is populated, and thereby do not consider
explicitly the pumping of the biexciton state $|i\rangle$. Further the excitonic level splitting  $\Delta = \omega_{\alpha}-\omega_{\beta}$ , $a_{0} =
(\gamma_{2}+\gamma_{4}+\gamma_{\alpha\beta}+\gamma_{\beta\alpha})$, $\Gamma_{a} =
(\gamma_{4}-\gamma_{2}+\gamma_{\alpha\beta}-\gamma_{\beta\alpha})$
and $A = \sqrt{\Gamma^{2}_{a}+4\gamma_{\alpha\beta}\gamma_{\beta\alpha}}$. 
The time dependent coefficients $C, D, F$ and $K$ are given by,\\
\begin{eqnarray}
\label{24}
C(t) &= &\left(2\gamma_{1}\left(1+\frac{\Gamma_{a}}{A}\right)+4\frac{\gamma_{3}\gamma_{\alpha\beta}}{A}\right)\frac{1-e^{-(a_{0}-A)t}}{2(a_{0}-A)}\nonumber\\
& &+ \left(A\rightarrow -A\right) ,\nonumber\\
D(t)&=&\left(2\gamma_{1}\left(1+\frac{\Gamma_{a}}{A}\right)+4\frac{\gamma_{3}\gamma_{\alpha\beta}}{A}\right)\frac{1-e^{-(a_{0}-A-2\gamma)t}}{2(a_{0}-A-2\gamma)}\nonumber\\
& &+ \left(A\rightarrow -A\right),\nonumber\\
F(t) &= &\left(2\gamma_{3}\left(1-\frac{\Gamma_{a}}{A}\right)+4\frac{\gamma_{1}\gamma_{\beta\alpha}}{A}\right)\frac{1-e^{-(a_{0}-A)t}}{2(a_{0}-A)}\nonumber\\
& &+ \left(A\rightarrow -A\right),\nonumber\\
K(t)&=&\left(2\gamma_{3}\left(1-\frac{\Gamma_{a}}{A}\right)+4\frac{\gamma_{1}\gamma_{\beta\alpha}}{A}\right)\frac{1-e^{-(a_{0}-A-2\gamma)t}}{2(a_{0}-A-2\gamma)}\nonumber\\
& &+ \left(A\rightarrow -A\right).\nonumber\\
\end{eqnarray}
The effect of non-degenaracy of the excitonic states and their incoherent dephasing on the dynamical evolution of the system shows up if one studies the two-time nonclassical second order correlation defined in equation (\ref{29}).
 For our four level system the explicit form of the
positive frequency part of the electric field operator is given by \cite{gsab},
\begin{eqnarray}
\label{28}
\vec{E}^{(+)}(\vec{r},t) & = &\vec{E}^{(+)}_{0}(\vec{r},t) -\left(\frac{\omega_{0}}{c}\right)\frac{1}{r}(\left[\hat{n}\times(\hat{n}\times\vec{d}_{\alpha i})\right]|\alpha\rangle\langle i|_{t}\nonumber\\
& + &\left[\hat{n}\times(\hat{n}\times\vec{d}_{\beta i})\right]|\beta\rangle\langle i|_{t}\nonumber\\
& + &\left[\hat{n}\times(\hat{n}\times\vec{d}_{j\alpha })\right]|j\rangle\langle \alpha|_{t}\nonumber\\
& + &\left[\hat{n}\times(\hat{n}\times\vec{d}_{j\beta})\right]|j\rangle\langle \beta|_{t}).
\end{eqnarray}
Finally using equation (\ref{28}) in (\ref{29}) we get the  general
form of the two time intensity-intensity correlation
\begin{eqnarray}
\label{30}
 \langle I I\rangle & = &\left(\frac{\omega_{0}}{c}\right)^{8}\frac{1}{r^4}\lbrace\langle[(\hat{\epsilon}_{H}\cdot\vec{d}_{\alpha i})^{\ast}\cos\theta_{1}|i\rangle\langle\alpha|_{t}\nonumber\\
 & + &(\hat{\epsilon}_{V}\cdot\vec{d}_{\beta i})^{\ast}e^{i\phi_{1}}\sin\theta_{1}|i\rangle\langle\beta|_{t}]\nonumber\\
&\times&(|\hat{\epsilon}_{H}\cdot\vec{d}_{j\alpha}|^{2}\cos^{2}\theta_{2}|\alpha\rangle\langle\alpha|_{t+\tau}+ |\hat{\epsilon}_{V}\cdot\vec{d}_{j\beta}|^{2}\sin^{2}\theta_{2}|\beta\rangle\langle\beta|_{t+\tau}\nonumber\\
& + &e^{-i\phi_{2}}(\hat{\epsilon}_{H}\cdot\vec{d}_{j\alpha})^{\ast}(\hat{\epsilon}_{V}\cdot\vec{d}_{j\beta})\cos\theta_{2}\sin\theta_{2}|\alpha\rangle\langle\beta|_{t+\tau}\nonumber\\
& + &e^{i\phi_{2}}(\hat{\epsilon}_{H}\cdot\vec{d}_{j\alpha})(\hat{\epsilon}_{V}\cdot\vec{d}_{j\beta})^{\ast}\cos\theta_{2}\sin\theta_{2}|\beta\rangle\langle\alpha|_{t+\tau})\nonumber\\
&\times&\left[(\hat{\epsilon}_{H}\cdot\vec{d}_{\alpha
i})\cos\theta_{1}|\alpha\rangle\langle
i|_{t}+(\hat{\epsilon}_{V}\cdot\vec{d}_{\beta
i})e^{-i\phi_{1}}\sin\theta_{1}|\beta\rangle\langle
i|_{t}\right]\rangle\rbrace.\nonumber\\
\end{eqnarray}
The two time correlation function that appears in equation
(\ref{30}) is evaluated by invoking the quantum
regression theorem \cite{lax} and equation(\ref{6}). Finally we get,
\begin{eqnarray}
\label{32}
\langle I I\rangle& = &\left(\frac{\omega_{0}}{c}\right)^{8}\frac{1}{4r^{4}}\mathcal{D}_{1}^{2}\mathcal{D}_{2}^{2}\langle|i\rangle\langle i|_{t}\rangle\nonumber\\
&\times&\lbrace f_{1}(\tau)+w_{1}(\tau)+f_{2}(\tau)+w_{2}(\tau)\nonumber\\
&+& (\cos2\theta_{1}+\cos2\theta_{2})(f_{1}(\tau)-w_{2}(\tau))\nonumber\\
&+&(\cos2\theta_{1}-\cos2\theta_{2})(w_{1}(\tau)-f_{2}(\tau))\nonumber\\
&+&\cos2\theta_{1}\cos2\theta_{2}(f_{1}(\tau)+w_{2}(\tau)-f_{2}(\tau)-w_{1}(\tau))\nonumber\\
&+& \sin2\theta_{1}\sin2\theta_{2}(e^{-i(\phi_{1}+\phi_{2})}u(\tau)+ e^{i(\phi_{1}+\phi_{2})}u^{\ast}(\tau))\rbrace.\nonumber\\
\end{eqnarray}
Here $\mathcal{D}_{1} = |\vec{d}_{\alpha j}| = |\vec{d}_{\beta j}|$ and $\mathcal{D}_{2} = |\vec{d}_{j \alpha}| = |\vec{d}_{j \beta}|$. The $f$'s, $w$'s and $u$ are found from the solutions of the density matrix equations (\ref{6}) and are given by,
\begin{eqnarray}
\label{33}
f_{1}(\tau) & = &e^{-a_{0}\tau}\left(\cosh(A\tau)+\frac{\Gamma_{a}}{A}\sinh(A\tau)\right) ,\nonumber\\
f_{2}(\tau) & = &2e^{-a_{0}\tau}\frac{\gamma_{\alpha\beta}}{A}\sinh(A\tau) ,\nonumber\\
w_{1}(\tau) & = &2e^{-a_{0}\tau}\frac{\gamma_{\beta\alpha}}{A}\sinh(A\tau) ,\nonumber\\
w_{2}(\tau) & = &e^{-a_{0}\tau}\left(\cosh(A\tau)-\frac{\Gamma_{a}}{A}\sinh(A\tau)\right) ,\nonumber\\
u(\tau) & = & e^{-(a_{0}-i\Delta)\tau}.
\end{eqnarray}
The equation (\ref{32}) gives the most general form of 
the two time intensity-intensity correlations for arbitrary
polarisation directions and for any system undergoing a cascade
emission. In the special case of $\phi_{1} = \phi_{2} = 0$, $\gamma_{\alpha\beta} = \gamma_{\beta\alpha}$ and $\gamma_{2} = \gamma_{4}$ this reduces to the simplified result (\ref{34}) of section (2).

\section{Conclusions}
In conclusion we have developed a simple theory to understand how the dephasing and energy level splitting of the excitonic states can affect polarisation entanglement of photons emitted in a biexciton-exciton cascade. We have also shown how these effects are important in determining whether the emitted photon pairs are classically correlated or entangled in different polarisation basis. Further we have shown that our theoretical calculation is in agreement with the experimental results found in context to such cascade emissions in quantum dots.  As a future prospect it would be interesting to extend the method of the present paper to a system of coupled dots or a dot in the micro-cavity.
\section{Acknowledgement}
This work was supported by NSF grant No. PHYS-0653494 and CCF-0829860.

\newpage

\end{document}